\documentclass{article}
\usepackage{amssymb}
\usepackage{amsmath,bm,shuffle}
\usepackage{array}
\usepackage{caption,graphicx}
\usepackage[all]{xy}

\newcommand{\be}{\begin{equation}}
\newcommand{\ee}{\end{equation}}

\begin{document}

\begin{flushright}
CERN-TH-2016-042
\end{flushright}

\vskip 1in

\begin{center}

{\Large Hyperlogarithms and periods in Feynman amplitudes}\footnote{Chapter 10 in:\textit{Springer Proceedings in Mathematics and Statistics}, \textbf{191}, V.K. Dobrev (ed.), 2016 ({\it Lie Theory and Its Applications to Physics} (LT-11), Varna, Bulgaria, 2015).}

\bigskip

{Ivan Todorov}

\medskip

{\small Institut des Hautes \'Etudes Scientifiques, 35 route de Chartres,
\\ F-91440 Bures-sur-Yvette, France;
\\ Theoretical Physics Department, CERN, CH-1211 Gen\`eve 23, Switzerland
\\ and
\\ Institute for Nuclear Research and Nuclear Energy, Bulgarian Academy of Sciences,
\\ Tsarigradsko Chaussee 72, BG-1784 Sofia, Bulgaria
\\ (permanent address)
\\e-mail address: ivbortodorov@gmail.com
}
\end{center}

\vglue 2cm

\begin{abstract}
The role of hyperlogarithms and multiple zeta values (and their generalizations) in Feynman amplitudes is being
gradually recognized since the mid 1990's. The present lecture provides a concise introduction to a fast developing
subject that attracts the interests of a wide range of specialists -- from number theorists to particle physicists.
\end{abstract}

\newpage

\tableofcontents

\bigskip


\newpage

\section*{Introduction}\label{sec0}
\addcontentsline{toc}{section}{Introduction}

Observable quantities in particle physics: scattering amplitudes, anomalous magnetic moments, are typically expressed in perturbation theory as (infinite) sums of {\it Feynman amplitudes} --  integrals over internal position or momentum variables corresponding to {\it Feynman graphs} (ordered by the number of internal vertices or by the number of loops -- the first Betti number of a graph). Whenever these integrals are divergent (which is often the case) one writes them as Laurent expansions in a (small) regularization parameter $\varepsilon$. (In the commonly used dimensional regularization $\varepsilon$ is half of the deviation of spacetime dimension from four: $2\varepsilon = 4 - D$. We shall encounter in Sect.~2 below a more general regularization with similar properties.) It was observed -- first as an unexpected curiosity in more advanced calculations (beyond one loop), then in a more systematic study -- that the resulting integrals involve interesting numbers like values of the zeta function at odd integers. Such numbers, first studied by Euler, but then forgotten for over two centuries, attracted independently, at about the same time the interest of mathematicians who defined the ${\mathbb Q}$-algebra ${\mathcal P}$ of {\it periods}. According to the elementary definition of Kontsevich and Zagier \cite{KZ} periods are complex numbers whose real and imaginary parts are given by absolutely convergent integrals of rational differential forms:
\be
\label{eq01}
I = \int_{\Sigma} \frac PQ \, dx_1 \ldots dx_n \qquad (\in {\mathcal P}),
\ee
where $P$ and $Q$ are polynomials with rational coefficients and the integration domain $\Sigma$ is given by polynomial inequalities again with rational coefficients.

\smallskip

Remarkably, the set of periods is denumerable -- they form a tiny (measure zero) part of the complex numbers but they suffice to answer all questions in particle physics. More precisely, it has been proven \cite{BW} that for rational ratios of invariants and masses all Laurent coefficients of dimensionally regularized euclidean Feynman amplitudes are periods. Brown \cite{B15} announces a similar result for convergent ``generalized Feynman amplitudes'' (that include the residues of primitively divergent graphs) without specifying the regularization procedure.

\smallskip

Amplitudes are, in general, functions of the external variables -- coordinates or momenta -- and of the masses of ``virtual particles'' associated with internal lines. Just like the numbers -- periods (that appear as special values of these functions) the resulting family of functions, the iterated integrals \cite{C}, has attracted independently the interest of mathematicians. Here belong the hyperlogarithms which possess a rich algebraic structure and appear in a large class of Feynman amplitudes, in particular, in conformally invariant massless theories.

\smallskip

The topic has become the subject of specialized conferences and research semesters\footnote{To cite a few: ``Loops and Legs in Quantum Field Theory'' Bi-annual Workshop taking place (since 2008) in various towns in Germany; Durham Workshop: ``Polylogarithms as a Bridge between Number Theory and Particle Physics'' \cite{Zh}; Research Trimester ``Multiple Zeta Values, Multiple Polylogarithms, and Quantum Field Theory'', ICMAT, Madrid, 2014, \cite{T16,V}.}. The present lecture is addressed, by contrast, to a mixed audience of mathematicians and theoretical physicists working in a variety of different domains. Its aim is to introduce the basic notions and to highlight some recent trends in the subject. We begin in Sect.~1 with a shortcut from the early Euler's work on zeta to the amazing appearance of his alternating (``$\phi$-function'') series in the calculation of the electron (anomalous) magnetic moment. Sect.~2 reviews the appearance of periods as residues of primitively divergent Feynman amplitudes. Sect.~3 introduces the double shuffle algebra of hyperlogarithms appearing inter alia in the calculation of position space conformal 4-point amplitudes. In Sect.~4 we introduce implicitly the formal multiple zeta values (MZV) defined by the double shuffle relations including ``divergent words'' and setting $\zeta (1) = 0$. The generating series $L(z)$ and $Z$ ``$= L(1)$'' are used to write down the monodromy around the possible singularities at $z=0$ and $z=1$ of the multipolylogarithms. We also display the periods of the ``zig-zag diagrams'' of Broadhurst and Kreimer and of the six-loop graph where a double zeta value $(\zeta (3,5))$ first appears. In Sect.~5 we define the ``multiple Deligne values'' (involving $N^{\rm th}$ roots of one) and provide a superficial glance at motivic zeta values \cite{BEK,B11,B12,B15} using them (following \cite{B11}) to derive the Zagier formula for the dimensions of the weight spaces of (motivic) MZVs. Finally, in Sect.~6 we give an outlook (and references to) items not treated in the text: single valued and elliptic hyperlogarithms and give, in particular, a glimps on the recent work of Francis Brown \cite{B15} that views the ``motivic Feynman periods'' as a representation of a ``cosmic Galois group'' revealing hidden structures of Feynman amplitudes.

\section{From Euler's alternating series to the electron magnetic moment}\label{sec1}
\setcounter{equation}{0}

Euler's interest in the zeta function and its alternating companion $\phi (s)$,
\be
\label{eq11}
\zeta (s) = \sum_{n=1}^{\infty} \frac1{n^s} \, , \quad \phi (s) = \sum_{n=1}^{\infty} \frac{(-1)^{n-1}}{n^s} \ \left( = (1-2^{1-s}) \, \zeta (s) \ \mbox{for} \ s > 1 \right)
\ee
was triggered by the {\it Basel problem} \cite{W} (posed by Pietro Mengoli in mid 17 century): to find a closed form expression for $\zeta (2)$. Euler discovered the non-trivial answer, $\zeta (2) = \frac{\pi^2}{6}$, in 1734 and ten years later found an expression for all $\zeta (2n)$, $n=1,2,\ldots$, as a rational multiple of $\pi^{2n}$. An elementary (Euler's style) derivation of the first few relations uses the expansion of $\cot (z)$ in simple poles (see \cite{B13}):
\begin{eqnarray}
\label{eq12}
z \cot (z) &= &1-2 \, z^2 \sum_{n=1}^{\infty} \ \frac{1}{n^2 \, \pi^2 - z^2} = 1-2 \sum_{n=1}^{\infty} \zeta (2n) \left(\frac{z^2}{\pi^2} \right)^n \nonumber \\
&= &\frac{1 - \frac{z^2}2 + \frac{z^4}{4!} - \ldots}{1 - \frac{z^2}{3!} + \frac{z^4}{5!} \ldots} \Longrightarrow \zeta (2) = \frac{\pi^2}{6} \, , \quad \zeta (4) = \frac{\pi^4}{90} , \ldots .
\end{eqnarray}
Euler tried to extend the result to odd zeta values but it did not work \cite{D12}. (We still have no proof that $\frac{\zeta (3)}{\pi^3}$ is irrational.) Trying to find polynomial relations among zeta values Euler was led by the {\it stuffle product}
\be
\label{eq13}
\zeta (m) \, \zeta (n) = \zeta (m,n) + \zeta (n,m) + \zeta (n+m)
\ee
to the concept of multiple zeta values (MZVs):
\be
\label{eq14}
\zeta (n_1 , \ldots , n_d) = \sum_{0 < k_1 < \ldots < k_d} \ \frac1{k_1^{n_1} \ldots k_d^{n_d}} .
\ee
The alternating series $\phi (s)$ (\ref{eq11}) (alias the Dirichlet {\it eta function}\footnote{Nowadays the term is usually associated with the {\it Dedekind $\eta$-function} $\eta (\tau) = e^{i \frac{\pi \tau}{12}} \overset{\infty}{\underset{n=1}{\prod}} \, (1-q^n)$, $q = e^{2\pi i \tau}$, defined on the upper half plane $\tau$.}) provide faster convergence in a larger domain. While $\zeta (s)$ has a pole for $s=1$, we have
\be
\label{eq15}
\phi (1) = \ln 2 .
\ee
Applying the stuffle relation for $\phi^2 (1)$:
\be
\label{eq16}
\phi^2 (1) (= (\ln 2)^2) = 2\phi (1,1) + \zeta (2)
\ee
where
\be
\label{eq17}
\phi (m,n) = \sum_{0 < k < \ell} \ \frac{(-1)^{k+\ell}}{k^m \ell^n} < 0
\ee
Euler expressed $\zeta (2)$ in terms of a much faster converging series (eventually guessing and then deriving
$\zeta(2)=\frac{\pi^2}{6}$ -- see Sect. 3 of \cite{Ay}):
$$
\zeta (2) = \phi (1)^2 - 2\phi (1,1) = (\ln 2)^2 + \sum_{n=1}^{\infty} \ \frac1{n^2 2^n}
$$
and computed it up to six digits $(\zeta (2) \approx 1.644934)$.

\smallskip

Remarkably, it is the same $\phi$-function which enters the $g$-factor of the magnetic moment of the electron ${\bm \mu} = g \, \frac{e}{2m} \, {\bm s}$ -- probably the most precisely measured and calculated quantity in physics \cite{K}. Up to third order in $\frac{\alpha}{\pi}$, where $\alpha = \frac{e^2}{4\pi \hbar c}$ is the fine structure constant, the anomalous magnetic moment $a_e = \frac{g-2}2$ is given by (\cite{LR,Sch}):
\begin{eqnarray}
\label{eq18}
a_e &= &\frac12 \, \frac{\alpha}{\pi} + \left[ \phi (3) - 6 \, \phi (1) \, \phi (2) + \phi (2) + \frac{197}{2^4 \, 3^2} \right] \left( \frac{\alpha}{\pi} \right)^2 \nonumber \\
&+ &\Biggl[ \frac2{3^2} \, (83 \, \phi (2) \, \phi (3) - 43 \, \phi (5)) - \frac{50}3 \, \phi (1,3) + \frac{13}5 \, \phi (2)^2  \\
&+ &\frac{278}3 \left( \frac{\phi (3)}{3^2} - 12 \, \phi \, (1) \, \phi (2) \right) + \frac{34202}{3^3 \, 5} \, \phi (2) + \frac{28259}{2^5 \, 3^4} \Biggl] \left( \frac{\alpha}{\pi} \right)^3 + \ldots . \nonumber
\end{eqnarray}
Schwinger's 1947 calculation of the first term $\left( \frac{\alpha}{2\pi} \right)$, contributing a $10^{-3}$ correction to Dirac's magnetic moment, won him (together with Tomonaga and Feynman) the 1965 Nobel Prize in physics. The second term (of order $\left( \frac{\alpha}{\pi}\right)^2$) was finally correctly calculated only 10 years later (by Peterman and Sommerfield). If Schwinger's work amounted to computing a single 1-loop (triangular) graph with 3 internal lines each, the 2-loop calculation involved 7 graphs, each new loop adding three additional lines (and as many new integrations in the Schwinger parameters -- see \cite{H04,K,St}). The three-loop calculation involving 72 graphs was completed first numerically (comparing with partial analytic results) by Kinoshita in 1995 and then fully analytically by Laporte and Remiddi a year later. The accuracy of both experimental measurement and theoretical computation (going nowadays beyond 4 loops!) is improving and the results match in a record of 12 significant digits (with uncertainty a part in a trillion):
\begin{eqnarray}
&&a_e = 1.159652 \, 180 \, 91 \, (\pm \, 26) \times 10^{-3} \ \mbox{(experiment)} \nonumber \\
&&a_e = 1.159652 \, 181 \, 13 \, (\pm \, 86) \times 10^{-3} \ \mbox{(theoretical).} \nonumber
\end{eqnarray}
In the words of ``a spectator'' in the ``tennis match between theory and experiment'' \cite{H04} ``20-year-long experiments are matched by 30-year-long calculations''. It is hard to overestimate the beauty and the significance of a formula like (\ref{eq18}) given the precision with which it is confirmed experimentally. The perturbative expansion is likely to be divergent but is believed to be asymptotic. Individual terms have a meaning of their own, both as special exactly known numbers and as measured quantity (the higher powers of $\frac{\alpha}{\pi}$ provide at least a hundred times smaller contribution). One is tempted to place these formulas among what Salviati (the \textit{alter ego} of Galileo) elevates to "those few which the human intellect does understand, I believe that its knowledge equals the Divine in objective certainty" \cite{G}

\smallskip

The $\phi$-function appearing in (\ref{eq18}) had a more than passing interest for Euler. In a 1740 paper (5 years after publishing his discovery of the formulas for $\zeta (2n)$, $n=1,2,\ldots ,6$) he wrote $\zeta (n) = N \pi^n$ indicating that for $n$-even, $N$ is rational while for $n$ odd he conjectures that $N$ is a function of $\ln 2$ (Sect.~6 of \cite{Ay}) -- a natural conjecture in view of (\ref{eq15}). In another paper of 1749 (of his Berlin period) after playing with some divergent series Euler proposes the functional equation for $\phi (s)$ writing: ``I shall hazard the following conjecture:
\be
\label{eq19}
\frac{\phi (1-s)}{\phi (s)} = - \frac{\Gamma (s) (2^s - 1) \cos \frac{\pi s}2}{(2^{s-1} - 1) \, \pi^s}
\ee
is true for all $s$.'' From  here and from (\ref{eq11}) the functional equation for $\zeta (s)$, proven by Riemann 110 years later (in 1859), follows immediately. Euler then admits that his earlier conjecture about odd zeta values went astray: ``I have already observed that $\phi (n)$ can only be computed for even $n$. When $n$ is odd all my efforts have been useless up to now.'' (Sect.~7 of \cite{Ay}.)

\section{Residues of primitively divergent amplitudes}\label{sec2}
\setcounter{equation}{0}

Let $\Gamma$ be a connected graph with finite sets ${\mathcal E}$ of {\it edges} (internal lines) and {\it vertices} ${\mathcal V}$, such that each edge $e \in {\mathcal E}$ is incident with a pair of different vertices ($v_i , v_j , v_i \ne v_j$ -- we do not allow for tadpoles). To each such graph we make correspond a position space Feynman integrand
$$
G_{\Gamma} (\vec x) = \prod_{e \, \in \, {\mathcal E}} G_e (x_{ij}) \, , \quad x_{ij} = x_i - x_j \ (x_i = (x_i^{\alpha} , \alpha = 1, \ldots , D))
$$
\be
\label{eq21}
1 \leq i,j \leq V \, , \quad \vec x = (x^1 , \ldots , x^N) \, , \quad N = D(V-1)
\ee
where $i,j$ label the vertices $v_i , v_j$, incident with the edge $e$, $V = \vert{\mathcal V}\vert$ is the number of vertices, $D$ is the spacetime dimension ($D = 4,6,\ldots$). We are in fact just interested in the case $D = 4$. Each {\it propagator} $G_e (x)$ is assumed to be locally integrable away from the origin. In the euclidean picture, to be used below (in which square intervals are given by $x^2 = \underset{\alpha}{\sum} (x^{\alpha})^2$) the integrands (\ref{eq21}) are actually smooth bounded (usually going to zero) at infinity functions away from the {\it large diagonal} ($x_i = x_j$ for some $i \ne j$). In Minkowski space $G_e$ is, generally, singular on the light cone $x^2 = 0$. The integrand (\ref{eq21}) is said to be {\it ultraviolet} (UV) {\it convergent} if it is locally integrable (at the diagonal) and hence gives rise to a (unique, tempered) distribution in ${\mathbb R}^N$. Otherwise, it is called (UV) divergent.

\smallskip

A (proper) subgraph $\gamma$ of $\Gamma$ is defined to contain a proper subset of vertices of $\Gamma$ together with the adjacent half edges and to contain every edge in $\Gamma$ incident with a pair of vertices $v_1 , v_2$ of $\gamma$. A divergent integrand $G_{\Gamma}$ is said to be {\it primitively divergent} if for any (connected) subgraph $\gamma \subset \Gamma$ the corresponding integrand $G_{\gamma}$ is convergent. In a massless quantum field theory (QFT) in which every propagator $G_e (x)$ is a rational homogeneous function of $x$,
\be
\label{eq22}
G_e (x) = \frac{p_e (x)}{(x^2)^{\mu_e}} \, , \quad \mu_e \in {\mathbb N} \, , \quad p_e (\lambda \, x) = \lambda^{\nu} p_e (x) \ \mbox{for} \ \lambda > 0
\ee
($\nu \leq 2 \, \mu_e$), there are simple convergence criteria in terms of homogeneity degrees \cite{NST}.

\bigskip

\noindent {\bf Proposition-Definition 2.1.} -- {\it A homogeneous density $G(\vec x) \, d^N x$ is convergent if its homogeneity degree is (strictly) positive. Otherwise, for
\be
\label{eq23}
G( \lambda \, \vec x) \, d^N \lambda \, x = \lambda^{-\kappa} \, G(x) \, d^N x \quad (d^N x = dx^1 \ldots dx^N) \, , \quad \kappa \geq 0
\ee
it is called} superficially divergent {\it with degree of superficial divergence $\kappa$.}

\bigskip

In a (massless) {\it scalar} QFT, in which all polynomials $p_e (x)$ are constants one proves that superficially divergent amplitudes are divergent. For more general spin tensor fields whose propagators have polynomial numerators a superficially divergent amplitude may, in fact, turn out to be convergent (see Sec.~5.2 of \cite{NST}). (Concering the role of Raymond Stora for the position space approach to renormalization - see [T].)

\smallskip

The following proposition (cf. \cite{NST}) serves as a definition of both the {\it residue} ${\rm Res} \, G$ and of renormalized primitively divergent amplitude $G^{\rho} (\vec x)$.

\bigskip

\noindent {\bf Proposition 2.2.} -- {\it If $G(\vec x)$ is primitively divergent then for any non-zero smooth seminorm $\rho(x)$ on ${\mathbb R}^N$ there exists a distribution ${\rm Res} \, G$ with support at the origin such that
\be
\label{eq24}
(\rho(\vec x))^{2\varepsilon} G(\vec x) - \frac1{\varepsilon} \, {\rm Res} \, G(\vec x) = G^{\rho} (\vec x) + O(\varepsilon)
\ee
where $G^{\rho} (\vec x)$ is a distribution valued extension of $G(\vec x)$ to ${\mathbb R}^N$. The calculation of the distribution ${\rm Res} \, G$ can be reduced to the case $\kappa = 0$ of a logarithmically divergent amplitude by using the identity
\be
\label{eq25}
({\rm Res} \, G)(\vec x) = \frac{(-1)^{\kappa}}{\kappa !} \, \partial_{i_1} \ldots \partial_{i_{\kappa}} {\rm Res} (x^{i_1} \ldots x^{i_{\kappa}} G (\vec x))
\ee
where summation is assumed (from $1$ to $N$) over the repeated indices $i_1 , \ldots , i_{\kappa}$. For a (reduced) $G$ that is homogeneous of degree $-N$ we have
\be
\label{eq26}
{\rm Res} \, G (\vec x) = {\rm res} \, G \, \delta (\vec x) \quad \mbox{ (whenever $\partial_i (x^i G) = 0$ for $x \ne 0$).}
\ee
Here the numerical residue ${\rm res} \, G$ is given by an integral over the (compact) projective space ${\mathbb P}^{N-1}$:
\be
\label{eq27}
{\rm res} \, G = \frac1{\pi^{N/2}} \int G (\vec x) \sum_{i=1}^N (-1)^{i-1} x^i \, dx^1 \wedge \ldots \wedge \widehat{dx^i} \wedge \ldots \wedge dx^N
\ee
(the hat over $dx^i$ means that it should be omitted).}

\bigskip

We note that for $D$ (and hence $N$) even $N-1$ is odd so that the space ${\mathbb P}^{N-1}$ is orientable.

\smallskip

Schnetz (Definition-Theorem 2.7 of \cite{Sch}) gives six equivalent expressions for the {\it ``period of a graph''} (i.e. for the residue of the corresponding amplitude). The statement, formulated for the massless $\varphi^4$ theory in $D=4$ is actually valid for any homogeneous of degree $-N$ ($= D (1-V)$ -- i.e. logarithmic) primitively divergent amplitude. In particular, the residue of a position space integrand $G (x_1 , \ldots , x_V)$ in a scalar QFT can be written as an $(N-D)$ dimensional integral
\be
\label{eq28}
{\rm res} \, G = \int \left( \prod_{i=2}^{V-1} \frac{d^D x_i}{\pi^{D/2}} \right) G(e,x_2,\ldots ,x_{V-1} , 0)
\ee
where $e$ is any ($D$-dimensional) unit vector $e^2 = 1$. For $D > 2$ the Schwinger parameter representation\footnote{See e.g. \cite{BS, PS} for a number theoretic application of this representation.} gives a still lower, $(L-1)$-dimensional, projective integral representation for the residue. (For a 4-point graph in the $\varphi^4$ theory the number of internal lines $L$ is related to the number of vertices $V$ by $L=2(V-1)$ so that $L-1 < D (V-2)$ for $D > 2$.)

\smallskip

In the important special case of a (massless) $\varphi^4$ theory in $D=4$ Schnetz \cite{Sch,S14} associates with each 4-point graph $\Gamma$ (i.e. a graph with 4 external half edges incident to four different vertices) a {\it completed $4$-regular vacuum graph} $\overline\Gamma$. He proves (Poposition 2.6 and Theorem 2.7 of \cite{Sch}) that all primitive 4-point graphs with the same primitive vacuum completion have the same residues. Moreover, there is a simple criterion allowing to tell when a 4-regular vacuum graph is primitive: namely, if the only way to split it by a four edge cut is by splitting off one vertex. (See examples in Sect. 4 below.)

\section{Conformal 4-point functions and hyperlogarithms}\label{sec3}
\setcounter{equation}{0}

Each primitively divergent 4-point Feynman amplitude in a (classically) conformally invariant QFT defines (upon integration over the internal vertices) a conformally covariant, locally integrable function away from the small diagonal $x_1 = \ldots = x_4$. On the other hand, every four points, $x_1 , \ldots , x_4$ can be confined by a conformal transformation to a 2-plane (by sending, say, a point to infinity and another to the origin). Then one can represent each {\it euclidean} point $x_i$ by a complex number $z_i$ so that
\be
\label{eq31}
x_{ij}^2 = \vert z_{ij} \vert^2 = (z_i - z_j)(\bar z_i - \bar z_j).
\ee
In order to make the map $x \to z$ explicit we fix a unit vector $e \in {\mathbb R}^4$ and let $n$ be a variable unit vector orthogonal to $e$ parametrizing a 2-sphere ${\mathbb S}^2$. Then any euclidean 4-vector $x$ can be written (in spherical coordinates) in the form
\be
\label{eq32}
x=r (\cos \rho \, e + \sin \rho \, n) \, , \quad e^2 = 1 = n^2 \, ,  \ en = 0 \, , \ r \geq 0 \, , \ 0 \leq \rho \leq \pi.
\ee
We make correspond to the vector (\ref{eq32}) a complex number $z$ such that
\be
\label{eq33}
z=r \, e^{i\rho} \to x^2 (=r^2) = z \bar z \, , \quad (x-e)^2 = \vert 1-z \vert^2 = (1-z)(1-\bar z) ;
\ee
\be
\label{eq34}
\int_{n \in \, {\mathbb S}^2} \frac{d^4 x}{\pi^2} = \vert z-\bar z\vert^2 \ \frac{d^2 z}\pi \left( \int_{{\mathbb S}^2} \delta (x) \, d^4 x = \delta (z) \, d^2z \right).
\ee
The 4-point amplitude with four distinct external vertices in the $\varphi^4$ theory has scale dimension 12 (in mass or inverse length units) and can be written in the form
\be
\label{eq35}
G(x_1 , \ldots , x_4) = \frac{f(u,v)}{\underset{i < j}{\prod} \, x_{ij}^2} = \frac{F(z)}{\underset{i < j}{\prod} \, \vert z_{ij} \vert^2}.
\ee
Here the (positive real) variables $u,v$; and the (complex) $z$ are conformally invariant cross ratios
\be
\label{eq36}
u = \frac{x_{12}^2 \, x_{34}^2}{x_{13}^2 \, x_{24}^2} = z \bar z \, , \quad v = \frac{x_{14}^2 \, x_{23}^2}{x_{13}^2 \, x_{24}^2} = \vert 1 - z \vert^2 \, , \quad z = \frac{z_{12} \, z_{34}}{z_{13} \, z_{24}} .
\ee

The cross ratios $z,\bar z$ are the simplest realizations of the argument of a hyperlogarithmic function whose graded Hopf algebra we proceed to define \cite{B,B09}.

\smallskip

Let $\sigma_0 = 0 , \sigma_1 , \ldots , \sigma_N$ be distinct complex numbers corresponding to an alphabet $X = \{ e_0 , \ldots , e_N\}$. Let $X^*$ be the set of (finite) words $w$ in this alphabet, including the empty word $\emptyset$. The hyperlogarithm $L_w(z)$ is an iterated integral \cite{C,B09} defined recursively in any simply connected open subset $U$ of the punctured complex plane $D = {\mathbb C} \backslash \Sigma$, $\Sigma = \{\sigma_0 , \ldots , \sigma_N\}$ by the {\it unipotent differential equations}\,\footnote{We use, following \cite{B11,S14}, concatenation to the right. Other authors \cite{B14,D} use the opposite convention.}
\be
\label{eq37}
\frac d{dz} \, L_{w\sigma} (z) = \frac{L_w (z)}{z-\sigma} \, , \quad \sigma \in \Sigma \, , \quad L_{\emptyset} = 1,
\ee
and the initial condition
\be
\label{eq38}
L_w (0) = 0 \ \mbox{for} \ w \ne 0^n = (\underbrace{0, \ldots , 0}_{n}) \, , \quad L_{0^n} = \frac{(\ln z)^n}{n!}.
\ee
There is a correpondence between iterated integrals and multiple power series; setting $n'_i = n_i - 1$, $k'_i = k_i - 1$ (and assuming $\sigma_i \ne 0$ for $1 \leq i \leq d$) we find
\be
\label{eq39}
(-1)^d \, L_{0^{n_0} \, \sigma_1 \, 0^{n'_1} \ldots \sigma_d \, 0^{n'_d}} (z) =
\ee
$$
\sum_{k_0 \geq 0 , k_i \geq n_i \, {\rm for} \, i = 1,\ldots , d \atop k_0 + \ldots + k_d = n_0 + \ldots + n_d} (-1)^{k_0 + n_0} \prod_{i=1}^d \begin{pmatrix} k'_i \\ n'_i \end{pmatrix} L_{0^{k_0}} (z) \, Li_{k_1 \ldots k_r} \left( \frac{\sigma_2}{\sigma_1} , \ldots , \frac{\sigma_d}{\sigma_{d-1}} , \frac z{\sigma_d} \right)
$$
where
\be
\label{eq310}
Li_{k_1 \ldots k_r} (z_1 , \ldots , z_r) = \sum_{0 < m_1 < \ldots < m_r} \frac{z_1^{m_1} \ldots z_r^{m_r}}{m_1^{k_1} \ldots m_r^{k_r}}.
\ee
The number of letters $\vert w \vert = n_0 + \ldots + n_d$ of a word $w$ defines its {\it weight}, while the number $d$ of non-zero letters is its {\it depth}. The product $L_w \, L_{w'}$ of two hyperlogarithms of weights $\vert w \vert , \vert w' \vert$ and depths $d,d'$ can be expanded in hyperlogarithms of weight $\vert w \vert + \vert w' \vert$ and depth $d+d'$, since the product of simplices can be expanded into a sum of higher dimensional simplices. In fact, the set $X^*$ of words can be equipped with a commutative {\it shuffle product} $w \shuffle w'$ defined recursively by
\be
\label{eq311}
\emptyset \shuffle w = w (= w \shuffle \emptyset), \quad au \shuffle bv = a (u \shuffle bv) + b(au \shuffle v)
\ee
where $u,v,w$ are (arbitrary) words while $a,b$ are letters (note that the empty word is {\it not} a letter). Denote by $O_{\Sigma}$ the ring of regular functions on $D$:
\be
\label{eq312}
O_{\Sigma} = {\mathbb C} \left[ z , \left( \frac1{z-\sigma_{\alpha}} \right)_{\alpha = 0,1,\ldots , N} \right].
\ee
Extending by $O_{\Sigma}$-linearity the correspondence $w \to L_w$ one proves that it defines a homomorphism of shuffle algebras $O_{\Sigma} \otimes {\mathbb C} (X) \to {\mathcal L}_{\Sigma}$ where ${\mathcal L}_{\Sigma}$ is the $O_{\Sigma}$ span of $L_w$, $w \in X^*$. The commutativity of the shuffle product is reflected in the identity
\be
\label{eq313}
L_{u \shuffle v} = L_u \, L_v \ (= L_v \, L_u).
\ee
If the shuffle relations are suggested by the expansion of products of iterated integrals, the product of series expansions of type (\ref{eq310}) suggests another commutative {\it stuffle product}. We illustrate the corresponding rule on the example of the product of a depth one and a depth two factors:
\begin{eqnarray}
\label{eq314}
Li_{n_1 , n_2} (z_1 , z_2) \, Li_{n_3} (z_3) &= &Li_{n_1 , n_2 , n_3} (z_1 , z_2 , z_3) + Li_{n_1 , n_3 , n_2} (z_1 , z_3 , z_2) \nonumber \\
&+ &Li_{n_3 , n_1 , n_2} (z_3 , z_1 , z_2) + Li_{n_1 , n_2 + n_3} (z_1 , z_2 z_3) \nonumber \\
&+ &Li_{n_1 + n_3 , n_2} (z_1 z_3 , z_2) .
\end{eqnarray}
(The corresponding product of words $u = z_1 \, 0^{n'_1} \, z_2 \, 0^{n'_2}$, $v = z_3 \, o^{n'_3}$ is denoted by $u * v$.) Clearly, the stuffle product also respects the weight but only filters the depth (there are terms of depth two and three in the right hand side of (\ref{eq314}) always not exceeding the total depth -- three -- of the left hand side). The shuffle and stuffle products give a number of relations among hyperlogarithms of the same weight. The monodromies of the multivalued hyperlogarithms around the possible singularities for $z = \sigma_{\alpha} \in \Sigma$ provide more (not easy to find) such relations. The bialgebra structure of hyperlogarithms introduced by Goncharov \cite{Gon} (see also Theorem 3.8 of \cite{B11} and Sect. 5.3 of \cite{D}) allow to reduce the calculation of monodromies and discontinuities  of higher weight hyperlogarithms to those of simple logarithms (see e.g. \cite{ABDG}). Here we shall just reproduce the coproduct for the special case of the classical polylogarithm:
\be
\label{eq315}
\Delta \, Li_n (z) = Li_n (z) \otimes 1 + \sum_{k=0}^{n-1} \ \frac{(\ln z)^k}{k!} \otimes Li_{n-k} (z) ,
\ee
the natural logarithm appearing as promitive element,
\be
\label{eq316}
\Delta \, L_{\sigma} (z) = L_{\sigma} (z) \otimes 1 + 1 \otimes L_{\sigma} (z) \, , \ L_{\sigma} (z) = \ln \left( 1-\frac{z}{\sigma} \right) = - Li_1 \left( \frac z{\sigma} \right)
\ee
$$
\mbox{for} \ \sigma \ne 0 \, , \quad L_0 (z) = \ln z .
$$
In order to apply (\ref{eq315}) to the specialization to $z=1$, $Li_n (1) = \zeta (n)$ for $n$ even we need to quotient the algebra of hyperlogarithms by $\zeta (2)$ or, better, by $\ln (-1) = i\pi \, (= \sqrt{-6 \, \zeta (2)})$ introducing the Hopf algebra ${\mathcal H}$:
\be
\label{eq317}
{\mathcal H} := {\mathcal L}_{\Sigma} / i\pi \, {\mathcal L}_{\Sigma} \quad \mbox{so that} \quad {\mathcal L}_{\Sigma} = {\mathcal H} [i\pi].
\ee
(Otherwise the relation $\zeta (4) = \frac25 \, \zeta^2 (2)$ would not be respected by the coproduct $\Delta$ satisfying, according to (\ref{eq315}), $\Delta \zeta (n) = \zeta (n) \otimes 1 + 1 \otimes \zeta (n)$.)

\smallskip

The coaction $\Delta$ is extended to ${\mathcal L}_{\Sigma}$ by
\be
\label{eq318}
\Delta : {\mathcal L}_{\Sigma} \to {\mathcal H} \otimes {\mathcal L}_{\Sigma} \, , \quad \Delta (i\pi) = 1 \otimes i\pi .
\ee
The asymmetry of the coproduct is reflected on its relation to differentiation and to discontinuity ${\rm dsic}_{\sigma} = M_{\sigma}-1$ (where $M_{\sigma}$ stands for the monodromy around $z=\sigma$):
\be
\label{eq319}
\Delta \left( \frac\partial{\partial z} \, F \right) = \left( \frac\partial{\partial z} \otimes {\rm id} \right) \Delta F \, , \quad \Delta ({\rm disc}_{\sigma} \, F) = ({\rm id} \otimes {\rm disc}_{\sigma}) \Delta F.
\ee
(One easily verifies, for instance, that both sides of (\ref{eq319}) give the same result for $F = Li_2 (z)$.) This allows us to consider ${\mathcal L}_{\Sigma}$ as a {\it differential graded Hopf algebra}.

\smallskip

The resulting structure allows to read off the symmetry properties of hyperlogarithms from the simpler properties of ordinary logarithms, as illustrated in Example 25 of \cite{D} which begins with a derivation of the inversion formula for the dilog:
$$
Li_2 \left( \frac1x \right) = i \pi \ln x - Li_2 (x) - \frac12 \ln^2 x + 2\zeta (2).
$$

\section{Multiple zeta values and Feynman periods}\label{sec4}
\setcounter{equation}{0}

The {\it multiple zeta values} (MZVs) are the values of the hyperlogarithms (\ref{eq310}) at arguments equal to one:
\be
\label{eq41}
\zeta (n_1 , \ldots , n_d) = Li_{n_1 \ldots n_d} (1,\ldots ,1) = (-1)^d \, \zeta_{10^{n'_1} \ldots 10^{n'_d}} \quad (n'_i = n_i - 1)
\ee
(cf. (\ref{eq39})). The corresponding series is convergent for $n_d > 1$. In order to recover the known relations among MZVs of the same weight one needs along with the shuffle and stuffle products of {\it convergent words} also a relation involving multiplication with the ``divergent word'' $e_1$ (in the case of a 2-letter alphabet, $\Sigma = \{0,1\}$, $X = \{e_0 , e_1 \}$):
\be
\label{eq42}
\zeta_{u \shuffle v} = \zeta_u \, \zeta_v = \zeta_{u * v} \, , \quad \zeta (e_1 \shuffle w - e_1 * w) = 0
\ee
for all convergent words $u,v$ and $w$. We note that the divergent words (with $n_d = 1$) in the last equation (\ref{eq42}) cancel out.  For instance, setting  $(n) = -e_1 \, e_0^{n-1}$, we find
\be
\label{eq43}
\zeta ((1) \shuffle (n) - (1) * (n)) = \sum_{i=1}^{n-1} \zeta (i , n+1-i) - \zeta (n+1) = 0 , \quad n \geq 2
\ee
a relation known to Euler. (Already for $n=2$ the resulting formula, $\zeta (1,2) = \zeta(3)$, is nontrivial.) Setting $(-\zeta_1 =) \zeta (1) = 0$ and using the relations (\ref{eq42}) one can write the {\it generating} series $Z$ of MZVs (also called {\it Drinfeld's associator}) in terms of multiple commutators of $e_0 , e_1$:
$$
Z = L(1) = 1 + \zeta (2) [e_0 , e_1] + \zeta (3) \left[ [e_0 , e_1] , e_0 + e_1 \right] + \ldots ,
$$
\be
\label{eq44}
\mbox{for} \ L(z) = 1 + \ln z \, e_0 + \ln (1-z) \, e_1 + \sum_{\vert w \vert \geq 2} L_w (z) \, w \quad (w \in X^* , X = \{ e_0 , e_1 \}).
\ee
(The limit of $L(z)$ for $z \to 1$ in the expression for $Z$ involves a regularization so that the divergent series for $-\zeta_1 = \underset{z \to 1}{\lim} \ \underset{n=1}{\overset{\infty}{\sum}} \frac{z^n}n$ is substituted by $0$.) The generating series (\ref{eq44}) allow to express in a compact form the monodromy of the multipolylogarithms $L_w (z)$ around the (possible) singularities at $z=0$ and $z=1$:
\begin{eqnarray}
\label{eq45}
{\mathcal M}_0 L(z) &= &e^{2\pi i e_0} L(z), \nonumber \\
{\mathcal M}_1 L(z) &= &Z \, e^{2\pi i e_1} Z^{-1} L(z).
\end{eqnarray}
(In writing $Z^{-1}$ we observe that any formal power series starting with 1 is invertible.)

\smallskip

There are infinitely many primitive vacuum graphs in the (massless) $\varphi^4$ theory.
They are completions of 4-point graphs with increasing number of loops $\ell = 3,4,\ldots$.
Their periods, up to $\ell = 6$, are MZVs of weight not exceeding $2\ell - 3$.
Broadhurst and Kreimer \cite{BK} discovered a remarkable sequence of {\it zig-zag graphs}
whose periods are rational multiples of $\zeta (2\ell - 3)$. Their completions are $n$-vertex
vacuum graphs $\overline\Gamma_n^{(2)} (n = \ell + 2)$ which admit a {\it hamiltonian cycle}
that passes through all vertices in consecutive order in such a way that each vertex $i$ is
also connected with $i \pm 2 \mod n$ (see Fig.~1a for the $n=8$ graph).
\begin{figure}[htb]
\centering
\includegraphics[width=0.7\textwidth]{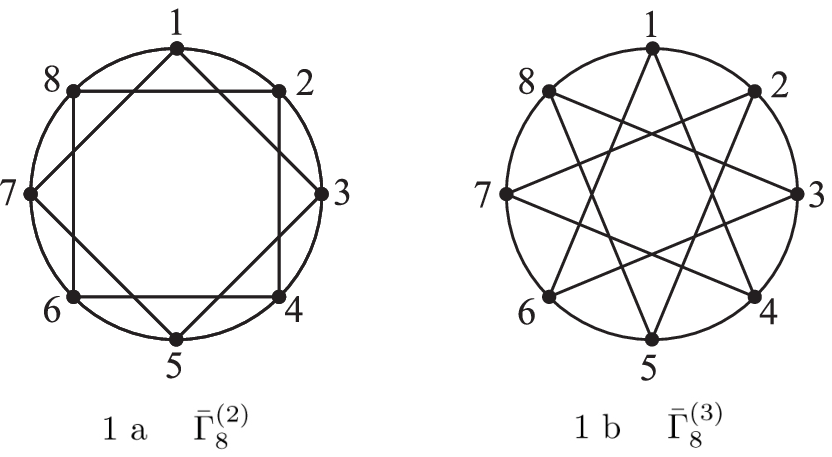}
\caption*{\footnotesize{Figure 1. Eight point vacuum completions of six-loop 4-point graphs in $\varphi^4 :{\rm Per} \,
(\overline\Gamma_8^{(2)}) = 24 \, \zeta (9)$, ${\rm Per} \, (\overline\Gamma_8^{(3)}) = 32 P_{3,5}$.}}
\label{Fig1}
\end{figure}

Their periods depend on the parity of $\ell$:
\begin{eqnarray}
\label{eq46}
{\rm Per} \, \left(\overline\Gamma_{\ell + 2}^{(2)}\right) &= &\frac{4-4^{3-\ell}}{\ell} \left({2 \ell - 2 \atop \ell - 1}\right) \zeta (2\ell - 3) \quad \mbox{for} \quad \ell = 3,5,\ldots \nonumber \\
&= &\frac4{\ell} \left({2 \ell - 2 \atop \ell - 1}\right) \zeta (2\ell - 3) \quad \mbox{for} \quad \ell = 4,6,\ldots
\end{eqnarray}
(a result conjectured in \cite{BK} and proven in \cite{BS12}). The 8-vertex graph on Fig.~1b also admits a hamiltonian cycle but with vertices $i$ connected with $i \pm 3 \mod 8$. Its period, computed numerically in \cite{BK}, is the first that involves a double zeta value:
\be
\label{eq47}
{\rm Per} \, \left(\overline\Gamma_8^{(2)}\right) = 32 \, P_{3,5} = 288 \left\{ \frac25 \, [29 \, \zeta (8) - 12 \, \zeta (3,5)] - 9 \, \zeta (3) \, \zeta (5) \right\}.
\ee
(The notation $P_{3,5}$ conforms with that of Brown \cite{B15}.) The first Feynman period, not expressible as a rational linear combination of MZVs was identified at 7 loops by E.~Panzer \cite{P} (following suggestions by Broadhurst and Schnetz) in 2014, as rational linear combination of hyperlogarithms at sixth roots of 1 (called {\it multiple Deligne values} in \cite{B14}). The 9-vertex vacuum completion of the graph in question is of type $F_9^{(3)}$: it again admits a hamiltonian cycle with hords joining vertices congruent mod 3 (as in the graph $\overline\Gamma_8^{(3)}$ displayed on Fig.~1b).

\section{Generalized and motivic MZVs}\label{sec5}
\setcounter{equation}{0}

Remarkably, MZVs $\zeta_w$ labeled by words in the $(N+1)$-point alphabet $X = \{ e_0 , e_1 , \ldots , e_N )$ corresponding to $\Sigma = \{ 0,1,\lambda , \ldots , \lambda^{N-1} \}$ where $\lambda$ is a primitive $N^{\rm th}$ root of unity again close a {\it double shuffle} (i.e. a shuffle and a stuffle) {\it algebra} and hence represent a natural generalization of the classical MZVs. In particular, the Euler $\phi$-function corresponds to $\Sigma = \{0,1,-1\}$:
$$
\phi (n) = L_{-10^{n'}} (1) = -Li_n (-1) \qquad (n' = n-1).
$$

Given the many relations these generalized MZVs $\zeta_w$ satisfy, the question arises to find a basis of such periods independent over the rationals. This question is wide open even for the classical MZVs (for which $w$ is a word in the two letter alphabet corresponding to $\Sigma = \{ 0,1\}$). We know the relations coming from (\ref{eq42}) but have no proof that there are no more relations for weights $\vert w \vert > 4$. If we denote by $d_n$ the dimension of the space of MZVs of weight $\vert w \vert = n$ we only know that $d_1 = 0$, $d_2 = d_3 = d_4 = 1$. (The reader is invited to verify -- using the relations (\ref{eq42}) (\ref{eq43}) -- that all MZVs of weight 4 are integer multiples of $\zeta (1,3) = \frac{\pi^4}{360}$ -- see Eq.~(B.8) of \cite{T14}.) We do not even have a proof that $\zeta (5)$ is irrational. A way to get around the resulting (difficult!) problem amounts to substitute the real MZVs by some abstract objects as {\it formal MZVs} (defined by the relations (\ref{eq42}) -- see \cite{S11,T16}) and {\it motivic zeta values} \cite{B11,B14} whose application for calculating the dimensions $d_n$ of the motivic MZVs of weight $n$ we proceed to sketch.

\smallskip

Consider the {\it concatenation algebra} ${\mathcal C}$ defined as the free algebra over the rational numbers ${\mathbb Q}$ on the countable alphabet $\{ f_3 , f_5 , \ldots \}$. The algebra of {\it motivic} MZVs is identified (non-canonically) with the algebra ${\mathcal C}[f_2]$ of polynomials in a single variable $f_2$ with coefficients in ${\mathcal C}$:
\be
\label{eq51}
{\mathcal C} [f_2] = {\mathcal C} \otimes_{\mathbb Q} {\mathbb Q} [f_2] , \quad {\mathcal C} = {\mathbb Q} \langle f_3 , f_5 , \ldots \rangle.
\ee
The algebra ${\mathcal C}[f_2]$ is graded by the weight (the sum of indices of $f_i$) and it is straightforward to compute the dimensions $d_n$ of the weight spaces ${\mathcal C}[f_2]_n$. Indeed, the generating {\it (Hilbert-Poincar\'e)} series for the dimensions $d_n^{\mathcal C}$ of the weight $n$ subspaces of ${\mathcal C}$ is given by
\be
\label{eq52}
\sum_{n \geq 0} d_n^{\mathcal C} \, t^n = \frac1{1-t^3 - t^5 - \ldots} = \frac{1-t^2}{1-t^2 - t^3},
\ee
while the generating series of ${\mathbb Q} [f_2]$ is $(1-t^2)^{-1}$. Multiplying the two series we obtain the dimensions $d_n$ of the weight spaces of (motivic) MZVs conjectured by Don Zagier:
\be
\label{eq53}
\sum_{n \geq 0} d_n \, t^n = \frac1{1-t^2 - t^3} \Rightarrow d_0 = 1  , d_1 = 0 ,  d_2 = 1 , d_{n+2} = d_n + d_{n-1} .
\ee

The concatenation algebra ${\mathcal C}$ can be equipped with a Hopf algebra structure (with $f_i$ as primitive elements) with the {\it deconcatenation coproduct} $\Delta : {\mathcal C} \to C \otimes C$ given by
\be
\label{eq54}
\Delta (f_{i_1} \ldots f_{i_r}) = 1 \otimes f_{i_1} \ldots f_{i_r} + f_{i_1 \ldots i_r} \otimes  1 + \sum_{k=1}^{r-1} f_{i_1} \ldots f_{i_k} \otimes f_{i_{k+1}} \ldots f_{i_r} .
\ee
The coproduct can be extended to the trivial comodule ${\mathcal C} [f_2]$ (\ref{eq51}) by setting
\be
\label{eq55}
\Delta {\mathcal C} [f_2] \to {\mathcal C} \otimes {\mathcal C} [f_2] \, , \quad \Delta (f_2) = 1 \otimes f_2
\ee
(and assuming that $f_2$ commutes with $f_{\rm odd}$). There exists a surjective {\it period map} of ${\mathcal C} [f_2]$ onto the space ${\mathcal Z}$ of real MZVs. The {\it main conjecture} in the theory of MZVs says that the period map is also injective, -- i.e., it defines an isomorphism of graded algebras. If true it would imply that the infinite sequence of numbers $\pi , \zeta (3) , \zeta (5), \ldots$ are transcendentals algebraically independent over the rationals \cite{Wa}. It would also give $\dim Z_n = d_n$. Presently, we only know that
\be
\label{eq56}
\dim Z_n \leq d_n \quad (\dim Z_n = d_n \ {\rm for} \ n \leq 4).
\ee

For weights $n \leq 7$ one can express all MZVs in terms of (products of) simple (depth 1) zeta values. For $n \geq 8$ this is no longer possible (as illustrated by the presence of $\zeta (3,5)$ in (\ref{eq47})). Brown \cite{B12} has established that the {\it Hoffman elements} $\zeta (n_1 , \ldots , n_d)$ with $n_i \in \{2,3\}$ form a basis of motivic zeta values for all weights (see also \cite{D12,Wa}).

\section{Outlook}\label{sec6}
\setcounter{equation}{0}

We shall sketch in this concluding section three complementary lines of development in the topic of our review.

\smallskip

The first goes in the direction of further specializing the class of functions and associated numbers (periods) appearing in the Feynman amplitudes of interest. Euclidean picture conformal amplitudes are {\it singlevalued} (as argued in \cite{GMSV,DDEHPS}). Knowing the monodromy (\ref{eq45}) one can construct a shuffle algebra of {\it single valued hyperlogarithms} \cite{B,B04} which belong to the tensor product $\overline{\mathcal L}_{\Sigma} \otimes {\mathcal L}_{\Sigma}$ of hyperlogarithms and their complex conjugates (see also \cite{T14,T16} for lightened reviews). The resulting functions and numbers, \cite{S14,B13}, are also encountered in superstring calculations (for a review see \cite{BSS}).

\smallskip

A second trend proceeds to considering massive Feynman amplitudes as well as higher order massless amplitudes which requires
extending the family of functions (and periods) of interest. The new functions appearing in the {\it sunrise} (or
{\it sunset}) graph in two and four dimensions are {\it elliptic hyperlogarithms} (for recent reviews and further references
-- see \cite{ABW,BKV,RT}). Modular forms and associated $L$-functions are also expected to play a role \cite{BS, B13, Br, PS}.

\smallskip

We do not touch another lively development championed by Goncharov and a group of physicists who also proceed to extending the mathematical tools -- using, in particular, {\it cluster algebras} -- in order to describe multileg amplitudes in $N=4$ supersymmetric Yang-Mills theory (for a review and references -- see \cite{V}).

\smallskip

A third approach attempts to reveal structures common to all Feynman amplitudes. Brown \cite{B15} gives a new meaning of the notion of {\it cosmic Galois group} (a term introduced by Cartier in 1998) {\it of motivic periods}: it is associated with the family of graphs with a fixed number of external lines and a fixed maximal number of different masses. Thus ${\mathcal C}_{4,0}$ is the cosmic Galois group associated with the 4-point amplitudes in a massless (say $\varphi^4$-) theory). Every Feynman amplitude of this class defines canonically a motivic period that gives rise to a finite dimensional representation of ${\mathcal C}_{4,0}$. One can associate a weight to motivic periods that generalizes the weight of MZVs. Brown proves that the space of motivic Feynman periods of a given type (say, the type $(4,0)$ above) of weight not exceeding $k$ is finite dimensional (Theorem 5.2 of \cite{B15}). This theorem allows to predict the type of periods of a given weight in amplitudes of any order. An illustration of what this means is the observation by Schnetz \cite{Sch} that the combination $\frac25 \, [29\zeta (8) - 12 \zeta (3,5)]$ of the period of the six loop graph corresponding to $\overline\Gamma_8^{(3)}$ (see (\ref{eq47})) also appears in a 7 loop period (multiplied by $252 \zeta (3)$ -- see Eq.~(6.2) of \cite{B15}). Brown remarks that the motivic version of the anomalous magnetic moment of the electron $a = \frac{g-2}2$ (Sect. I) also displays some compatibility with the action of the cosmic Galois group on periods.

\vglue 1cm

\noindent {\bf Acknowledgments.} It is a pleasure to thank V. Dobrev, E. Ivanov, and G. Zoupanos for their invitations and hospitality to  Varna, Bulgaria, Dubna, Russia and Corfu, Greece, respectively, where I was given the opportunity to talk on and discuss the topic of this paper. I thank IHES and the Theoretical Physics Department of CERN for hospitality during the course of this work. The author's work is supported in part by Grant DFNI T02/6 of the Bulgarian National Science Foundation.

\end{document}